\documentclass[prl,aps,floatfix,twocolumn,superscriptaddress]{revtex4}

\usepackage{graphicx}
\graphicspath{{figures/}}

\usepackage{amsmath}
\usepackage{units,xspace}
\newcommand{\micron}{\ensuremath{\unit{\mu m}}\xspace}

\newcommand{\abs}[1]{\left\vert #1 \right\vert}
\newcommand{\avg}[1]{\left< #1 \right>}
\renewcommand{\vec}[1]{\ensuremath{\boldsymbol{#1}}}
\newcommand{\uvec}[1]{\ensuremath{\hat{\boldsymbol{#1}}}}

\begin{document}

\title{Influence of non-conservative optical forces on the
dynamics of optically trapped colloidal spheres: 
The fountain of probability}

\author{Yohai Roichman}
\affiliation{Department of Physics and Center for Soft Matter
  Research, New York University, New York, NY 10003}

\author{Bo Sun}
\affiliation{Department of Physics and Center for Soft Matter
  Research, New York University, New York, NY 10003}

\author{Allan Stolarski}
\affiliation{{NEST+m}, New York, NY 10002}

\author{David G. Grier}
\affiliation{Department of Physics and Center for Soft Matter
  Research, New York University, New York, NY 10003}

\date{\today}

\begin{abstract}
We demonstrate both experimentally and theoretically that
a colloidal sphere trapped in a static optical tweezer
does not come to equilibrium, but rather reaches a steady state
in which its probability flux traces out a toroidal vortex.
This non-equilibrium behavior can be ascribed to a subtle bias
of thermal fluctuations by non-conservative optical forces.
The circulating 
sphere therefore acts as a Brownian motor.
We briefly discuss ramifications of this effect for
studies in which optical tweezers have been
treated as potential energy wells.
\end{abstract}

\pacs{82.70.Dd, 87.80.Cc}

\maketitle

Most discussions of the dynamics of optically trapped particles
assume at least implicitly that the forces exerted by an
optical tweezer \cite{ashkin86} are path-independent and
therefore conserve mechanical energy.
Optical forces due to gradients in the intensity are
manifestly conservative in this sense \cite{ashkin92}.
Radiation pressure, by contrast, is not
\cite{ashkin92,roichman08}.
The experimental studies described in this Letter
demonstrate that the non-conservative component of the
optical force has measurable consequences for the 
dynamics of optically trapped colloidal spheres.
In particular, the probability density
for a sphere trapped in a static optical tweezer
exhibits steady-state toroidal currents,
a phenomenon we call a fountain of probability.
We use the Fokker-Planck formalism to explain how
non-conservative forces bias random thermal fluctuations
to induce circulating probability currents.

\begin{figure}
  \centering
  \includegraphics[width=\columnwidth]{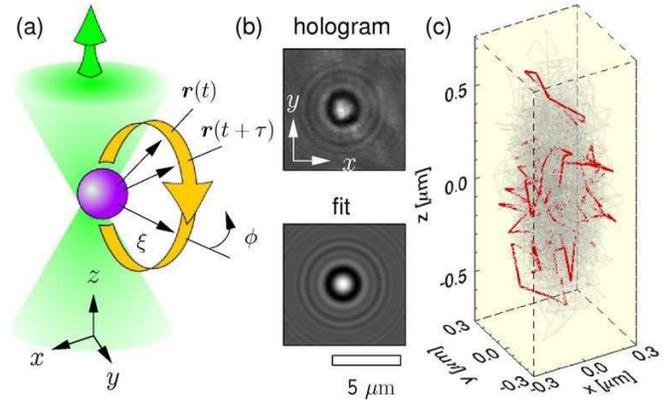}
  \caption{(Color online) (a) Schematic representation of a colloidal sphere 
    in an optical tweezer created from a beam of light propagating
    along $\uvec{z}$.  The curving arrow represents a thermally
    driven trajectory.  (b) Measured
    hologram of a trapped sphere together with a nonlinear least-squares
    fit to Mie scattering theory. (c) Measured 5~\unit{min}
    trajectory in a trap at $P_0 = 7.5~\unit{mW}$.  Emphasized portion
    shows 10~\unit{s}.}
  \label{fig:fount}
\end{figure}

Figure~\ref{fig:fount} schematically represents
the deceptively simple system.
A single colloidal sphere is drawn to the focus
of a converging laser beam by forces
arising from gradients in the beam's intensity
\cite{ashkin86,ashkin92}.
These intensity-gradient forces establish
a three-dimensional potential energy well,
$V(\vec{r})$, determined
by the local intensity, $I(\vec{r})$.
A particle trapped in this well
also experiences radiation pressure that
drives it downstream with a force 
proportional to $I(\vec{r})$.
In the absence of thermal fluctuations, a trapped
particle would come to rest at 
a stable mechanical equilibrium
downstream of the focus.

Treating the displaced equilibrium point as the
origin of an effective potential energy well
is tempting but misleading.
To appreciate the problem,
consider a thermally driven trajectory
such as the example shown schematically in
Fig.~\ref{fig:fount}.
Were the system in thermodynamic equilibrium, forward and
reverse trajectories around this loop would
have equal probability.
Because the light is more intense near the optical axis, however,
radiation pressure biases the random walk in
the forward direction.
This departure from detailed balance should
induce irreversible circulation in the particle's
otherwise random fluctuations \cite{tomita74}.

We demonstrate this effect by observing the motions of
colloidal silica spheres
2.2~\micron in diameter (Bangs Laboratories, SS04N/7651)
dispersed in water and individually trapped 
in discrete optical tweezers.
Each trap is formed from roughly 1~\unit{mW} of laser light at a vacuum
wavelength of 532~\unit{nm} (Coherent Verdi) projected with the
holographic optical trapping technique
\cite{dufresne98,grier03,polin05}.
The beams are brought to a diffraction-limited focus
by an objective lens (Nikon Plan Apo $100\times$, oil immersion)
mounted in an inverted optical microscope (Nikon TE-2000U).
We track the particle in three dimensions
with nanometer precision using 
video holographic microscopy
\cite{sheng06,lee07,lee07a}.
In-line holograms, such as
the example in Fig.~\ref{fig:fount}(b), are created
by illuminating the sphere with a collimated laser beam,
in this case the 3~\unit{mm} diameter beam provided by a
10~\unit{mW} He-Ne laser (Uniphase) operating at a vacuum
wavelength of 633~\unit{nm}.
Light scattered by the particle interferes with
the unscattered portion of the incident beam to
produce an interference pattern in the microscope's focal plane.
This is magnified and projected onto a CCD camera
(NEC TI-324AII), which records holograms at video rates.
Each frame in the video stream then is fit to the predictions
of Lorenz-Mie scattering theory \cite{bohren83}
to obtain each particle's position in three dimensions,
$\vec{r}(t)$,
its radius $a$, and its complex index of refraction
\cite{lee07a}.
Figure~\ref{fig:fount}(b) also shows the computed
image resulting from fitting to the experimental image.
A sequence of such fits yields the particle's 
three-dimensional trajectory,
as plotted in Fig.~\ref{fig:fount}(c).

Were the particle to come to equilibrium within the trap,
the probability to find it within
$d\vec{r}$ of $\vec{r}$ would be given by the
Boltzmann distribution,
\begin{equation}
  \label{eq:boltzmann}
  \rho(\vec{r}) = N \, \exp\left(- \beta V(\vec{r})\right),
\end{equation}
where $\beta^{-1} = k_B T$ is the thermal energy scale at
absolute temperature $T$ and $N$ is a normalization constant.
This is the basis for a popular method to calibrate optical
traps \cite{florin98},
with measured particle positions being compiled
into $P(\vec{r})$, which then is inverted to yield
$V(\vec{r})$.
In the further approximation that the trap can be
modeled as a three-dimensional
harmonic well centered at the origin, the three-dimensional
probability density can be factored into one-dimensional
contributions, 
$\rho(\vec{r}) = \prod_{j=1}^3 N_j \, 
\exp\left( - \beta V_j(r_j) \right)$,
where
\begin{equation}
  V_j(r_j) = \frac{1}{2} \, k_j r_j^2.
  \label{eq:harmonic}
\end{equation}
The trap is then characterized by an effective spring constant
$k_j$
in each of the three Cartesian coordinates.

\begin{figure}[!t]
  \centering
  \includegraphics[width=\columnwidth]{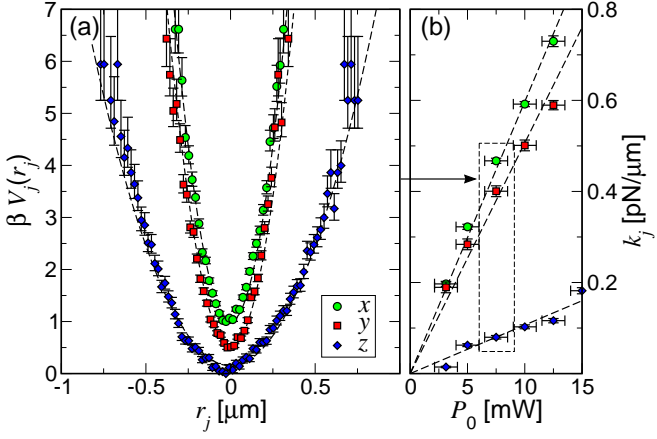}
  \caption{(Color online) (a)
    Effective potential localizing a colloidal sphere to 
    an optical trap at laser power $P_0 = 7.5~\unit{mW}$, 
    in units of the thermal energy scale $k_B T$.  
    Dashed curves are parabolas whose curvatures yield the trap
    stiffness.  Discrete points
    are obtained from histograms, $P_j(r_j)$, of the measured particle
    positions in 20~\unit{nm} bins.  The $x$ and $y$ results are
    offset by $1~k_B T$ and $0.5~k_B T$, respectively, for clarity.
    (b) Dependence of the effective trap stiffness on laser power.}
  \label{fig:well}
\end{figure}

Figure~\ref{fig:well}(a) shows the result of
applying Eq.~(\ref{eq:harmonic}) to
10,000
measurements of a single particle's position in an
optical tweezer powered by $P_0 = 7.5 \pm 1.0~\unit{mW}$
of light.
The difference between the trap's apparent stiffness
$k_x = 0.467 \pm 0.009~\unit{pN/\micron}$
along the axis of the light's polarization and
$k_y = 0.400 \pm 0.008~\unit{pN/\micron}$ in the
perpendicular direction is consistent with previous
reports of polarization effects in optical trapping
\cite{rohrbach05,zakharian06,dutra07}.
The axial stiffness, $k_z = 0.080 \pm 0.004~\unit{pN/\micron}$
is a factor of 5 smaller because of the comparatively weak
axial intensity gradients \cite{vermeulen06}.
The dependence on laser power plotted in Fig.~\ref{fig:well}(b)
confirms that $k_j \propto P_0$.

\begin{figure}[!t]
  \centering
  \includegraphics[width=\columnwidth]{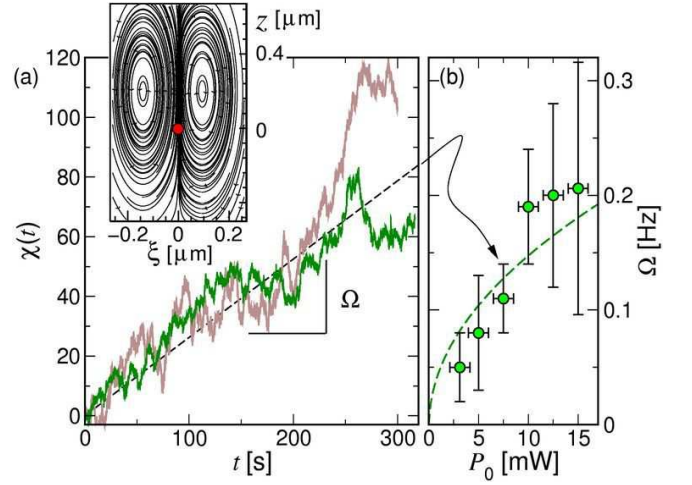}
  \caption{(Color online) (a) Accumulated circulation of a single particle's 
    trajectory over 10,000 time steps at laser power 
    $P_0 = 7.5~\unit{mW}$.  Dark (green) trace: experimental data.
    Light (brown) trace: simulation results for equivalent
    conditions.
    Inset: streamlines of the measured velocity field in the
    $(\xi,z)$ plane, with the trap's center denoted by a circle.
    (b) Circulation rate as a function of laser power together with
    prediction of Eq.~(\ref{eq:circulation}).
     }
  \label{fig:chi}
\end{figure}

To quantify nonequilibrium effects in the particle's dynamics,
we consider its trajectory in cylindrical coordinates,
$\vec{r} = (\xi, \phi, z)$, centered
on the trap's equilibrium position.
During its thermally-driven exploration of the trap, the
particle covers a roughly elliptical region
in the $(\xi,z)$ plane.
A statistical bias due to radiation pressure
along the axis
should appear as a tendency of
the particle's trajectory to wind clockwise in the
$(\xi, z)$ plane.
To quantify this, we define a measure of the trajectory's
circulation over the $\tau = 1/30~\unit{s}$
the interval between video frames,
\begin{equation}
  \label{eq:areafraction}
  \Omega(t) \equiv 
  \frac{1}{2\pi} \,
  \frac{ (\vec{r}(t+\tau) \times \vec{r}(t)) \cdot \uvec{\phi}}{
    \sqrt{\avg{(\xi - \avg{\xi})^2} \, 
      \avg{(z - \avg{z})^2}}},
\end{equation}
which is positive for clockwise circulation and negative for
retrograde motions.
The angle brackets in Eq.~(\ref{eq:areafraction})
denote an average over the trajectory.
To identify trends in $\Omega(t)$ despite large thermal
fluctuations, we further define
\begin{equation}
  \label{eq:chi}
  \chi(t) \equiv \int_0^t \Omega(t^\prime) \, dt^\prime
\end{equation}
which measures the accumulation of clockwise circulation in
the particle's trajectory.
The increasing trend in $\chi(t)$
confirms the existence of a
circulating steady state in which the
particle completes one net cycle in roughly 9~\unit{s}
with a mean drift speed of roughly
$150~\unit{nm/s}$.

The measured circulation rate increases with laser
power, as shown in Fig.~\ref{fig:chi}(b).
The particle also becomes increasingly well localized
as the trap stiffens, however, so that 
the range of experimentally accessible laser powers
is limited to $P_0 < 15~\unit{mW}$ 
by our instrumental resolution.



Observing a convective flux in the trapped 
particles' trajectories confirms the system's
departure from equilibrium.
The nature of the nonequilibrium state is clarified
by considering an idealized model of the system.
The probability flux induced by a force $\vec{F}(\vec{r})$
acting on a Brownian particle is
\begin{equation}
  \vec{S}(\vec{r}) = \mu \rho \vec{F} + D \nabla \rho,
\end{equation}
where $\rho(\vec{r})$ is the ensemble-averaged
probability density for finding the particle near $\vec{r}$.
We assume that the system reaches steady state so that
$\rho(\vec{r})$ does not depend on time.
Continuity of the probability density 
($\nabla \cdot \vec{S} = 0$) then
yields the Fokker-Planck equation
\begin{equation}
  D \nabla^2 \rho = \mu \vec{F} \cdot \nabla \rho
  + \mu \rho \nabla \cdot \vec{F}.
  \label{eq:fokkerplanck}
\end{equation}
The divergence-free flux still
can support circulation,
\begin{equation}
  \nabla \times \vec{S} = \mu \vec{F} \times \nabla \rho + 
  \mu \rho \nabla \times \vec{F}
\end{equation}
provided $\nabla \times \vec{F} \neq 0$.

We model the trap as a radially symmetric harmonic well
with radiation pressure directed along the optical axis
with a strength proportional to the local
intensity:
\begin{equation}
  \label{eq:trap}
  \vec{F}(\vec{r}) 
  = 
  - k \vec{r}
  + f_1 \, \exp\left(- \frac{r^2}{2 \sigma^2} \right) \, \uvec{z}.
\end{equation}
This model is parametrized the trap's 
stiffness, $k$, and the 
scale of radiation pressure, $f_1$, both of
which are proportional to the laser's power, $P_0$.
and also by
the effective width of the trap, $\sigma$, which depends on
the quality of the focus and the size of the trapped particle
\cite{ladavac04,pelton04a,roichman07a}.
For a particle whose radius $a$ is substantially
larger than the wavelength of light, $\sigma \gtrsim a$.
Stable trapping requires
the restoring force to exceed radiation pressure, so that
$\epsilon = f_1/(k\sigma)$ may be taken to be small
parameter independent of laser power.
In that case, $\vec{F}(\vec{r})$ has a point of stable
mechanical equilibrium at $z = \epsilon \sigma$.

The lighter trace in Fig.~\ref{fig:chi}(a) shows the 
circulation observed in
a fourth-order Runge-Kutta 
Brownian dynamics simulation of a particle diffusing
in the force field described by Eq.~(\ref{eq:trap}), using
parameters obtained from Fig.~\ref{fig:well}, and assuming
$\epsilon \approx 0.1$.
Both the slope of $\chi(t)$ and the magnitude of its
fluctuations are consistent with experimental results.
This agreement both supports our
interpretation of our experimental observations and
also indicates that Eq.~(\ref{eq:trap}) is sufficiently
detailed for quantitative comparisons with our measurements.

Expanding the probability density to first order in $\epsilon$,
$\rho(\vec{r}) = \rho_0(r) + \epsilon \rho_1(\vec{r})$,
yields an approximate solution 
\cite{fountainepaps} to Eq.~(\ref{eq:fokkerplanck}),
\begin{align}
  \label{eq:rho0}
  \rho_0(r) 
  & = 
  \left(\frac{\beta k}{2 \pi}\right)^{\frac{3}{2}} \,
  \exp\left(-\frac{1}{2} \beta k \, r^2 \right) \quad \text{and}\\
  \label{eq:rho1}
  \rho_1(\vec{r}) 
  & \approx 
  \beta k \sigma \, 
  \exp\left(- \frac{r^2}{2 \sigma^2} \right) \, z \, \rho_0(r).
\end{align}
Equation~(\ref{eq:rho0}) is the result that would be obtained for
a particle at equilibrium in a harmonic potential.
Equation~(\ref{eq:rho1}) describes the lowest-order correction
due to spatially nonuniform radiation pressure.
Because this expression for $\rho_1(\vec{r})$ is obtained in
the stable-trapping limit, $\beta k \sigma^2 \gg 1$, it
underestimates the probability
for the particle to make large-scale excursions from the trap.
The results that follow therefore are
conservative underestimates for the influence of 
non-conservative optical forces on a trapped particle's dynamics.

Even if the particle were to reach equilibrium in the force
field $\vec{F}(\vec{r})$, the distortion of $\rho_0(r)$ by
$\rho_1(\vec{r})$ would affect measurements of colloidal
forces calibrated by thermal fluctuation analysis.
The first-order correction not only
displaces the center of the probability distribution downstream, but
also broadens it.
Consequently, the effective trap stiffness estimated
from a probe particle's unloaded fluctuations will
systematically underestimate the forces exerted by the trap
under load, when fluctuations are suppressed.

Our experimental results demonstrate, furthermore, that
the distorted probability density does not come to equilibrium,
but rather undergoes
steady-state circulation.
Expanding $\vec{S}(\vec{r})$ to lowest non-vanishing
order in $\epsilon$,
we obtain the mean circulation rate \cite{fountainepaps}
\begin{align}
  \Omega_0 
  & = 
  \frac{1}{2\pi} \, \int \left(\nabla \times \vec{S}\right) \cdot \uvec{\phi} \, d^3r \\
  & \approx 
 \frac{1}{\sqrt{18\pi}} \, \frac{\mu f_1}{\sigma} \, 
  \frac{(\beta k \sigma^2)^{\frac{1}{2}}}{1 + \beta k \sigma^2}.
  \label{eq:circulation}
\end{align}

This result can be generalized for an anisotropic trap by
appropriately scaling the Cartesian coordinates in
Eq.~(\ref{eq:trap}).
In that case, Eq.~(\ref{eq:circulation}) still holds with
the effective spring constant set by the harmonic mean,
$k = 3 / (k_x^{-1} + k_y^{-1} + k_z^{-1})$.
Given the results from Fig.~\ref{fig:well}, 
the measured circulation rates in Fig.~\ref{fig:chi}(b) agree
well with the predictions of
Eq.~(\ref{eq:circulation}) for $\epsilon = 0.11 \pm 0.04$,
including the predicted $P_0^{\frac{1}{2}}$ dependence on laser power.

Equation~(\ref{eq:circulation}) also predicts that the
circulation rate scales with temperature as
$\Omega_0 \sim T^{\frac{1}{2}}$ for $\beta k \sigma^2 > 1$.
This confirms that the particle
would not circulate at all were it not for thermal fluctuations.
The optically trapped particle therefore is an exceptionally
simple example of a Brownian motor
\cite{reimann96,reimann02,reimann02b} 
whose ability to perform work
relies on rectification of thermal noise.
Unlike previous optical implementations of thermal ratchets
\cite{faucheux95,faucheux95a,lee05,lee05a,lee06},
this fountain of probability requires no time-dependent driving,
but rather is biased into motion by the
non-conservative component of the optical force \cite{roichman08}.

The fountain's efficiency 
as a Brownian motor can be estimated by computing
the power dissipated into the water by viscous drag given
the particle's mean drift velocity, 
$\vec{v}(\vec{r}) = \vec{S}(\vec{r})/\rho(\vec{r})$
\cite{seifert05}.
The result \cite{fountainepaps},
\begin{equation}
  \label{eq:power}
  P 
  =
  \int \frac{\abs{\vec{S}}^2}{\mu \, \rho} \, d^3r
  \approx
  5 \mu f_1^2 \, \frac{(\beta k \sigma^2)^{\frac{3}{2}}}{
    (2 + \beta k \sigma^2)^{\frac{7}{2}}}
\end{equation}
depends strongly on laser power at low powers but becomes
independent of $P_0$ in the strong-trapping limit.
The maximum dissipation rate,
$P_\text{max} = 5 \mu \epsilon^2 (k_B T)^2 / \sigma^2$,
is a fixed proportion of the particle's thermally-driven 
velocity fluctuations.
The system transduces fluctuations into heat through
the bias imposed by the non-conservative
optical force.
Its efficiency is limited by the material properties that
determine $\epsilon$, and thus by the requirement that
the particle remain trapped.

Although this model is consistent with our measurements,
other mechanisms also might cause a particle to
circulate in an optical trap.  For example,
localized heating due to absorption of light
at the focal point could create a toroidal convection roll
that would advect the particle.
This has been demonstrated with infrared optical tweezers
operating at 1480~\unit{nm} \cite{braun02}.
The optical absorption coefficient of water
is five orders of magnitude smaller at 532~\unit{nm},
however \cite{hale73}.
Scaling the previously reported \cite{braun02}
convection rate accordingly suggests that
the maximum drift speed due to thermal convection in our system 
should be no greater than $1~\unit{nm/s}$ at the highest
laser powers in out study.
Thermal convection therefore appears unlikely to account for 
our observations.

If, indeed, steady-state circulation is an intrinsic feature
of optically trapped particles' motions,
the nonequilibrium effects we have identified
should influence any measurement based on analysis of their thermal
fluctuations.
How such effects might have affected previously reported
measurements remains to be determined.
Establishing that a colloidal particle trapped in
a static optical tweezer acts as a
Brownian motor also creates new opportunities for research
in nonequilibrium statistical physics.
For example, the nonuniform intensity in an optical tweezer
also exert position-dependent torques on the trapped particle
that we have not considered here.
The resulting nonequilibrium behavior in the particle's
rotational degrees of freedom may share features in common
with the fountain-like translational bias we have identified,
and the two may be coupled in interesting ways.
These considerations also may be extended to account for
nonequilibrium effects in more general light fields.

We are grateful to Vincent Pereira and Alexander Grosberg
for enlightening conversations.
This work was supported by the National Science Foundation
through Grant Number DMR-0451589 and by a grant from
Consolidated Edison.


\end{document}